# Does the momentum flux generated by gravitational contraction drive AGB mass-loss?

B. M. Lewis

Arecibo Observatory

Arecibo Observatory,
PO Box 995,
Arecibo PR 00613

**Abstract:**

Gravitational contraction always generates a radially directed momentum flux. A particularly simple example occurs in the electron-degenerate cores of AGB stars, which contract steadily under the addition of helium ashes from shell hydrogen burning. The resulting momentum flux is quantified here. And since the cores of AGB stars lack efficient momentum cancellation mechanisms, they can maintain equilibrium by exporting their excess momentum flux to the stellar envelope, which disposes of much of it in a low velocity wind. Gravitational contraction easily accounts for the momentum flux in the **solar wind**, as well as the flux required to lift mass into the dust formation zone of every AGB star, whereon radiation pressure continues its ejection as a low velocity wind. This mechanism explains the dependence of the AGB mass-loss rate on core mass; its generalization to objects with angular momentum and/or strong magnetic fields suggests a novel explanation of why most planetary nebulae and proto planetary nebulae exhibit axial symmetry.

Quasistatic contraction is inherently biased to the generation of the maximum possible momentum flux. Its formalism is therefore readily adapted to providing an upper limit to the momentum flux needed to sustain mass loss when this begins from a semi-continuous rather than impulsive process.

Gravitation
Stars: Mass-Loss
Stars: AGB & post-AGB
Sun: Solar Wind

I. Introduction

Mass loss from the near environs of self gravitating bodies is ubiquitous. It occurs in spectacular fashion in the bipolar jets emanating from AGN such as M82, NGC4258 & M87, from quasars such as 3C273, as well as from stellar black holes such as SS433. Bipolar flows frequently occur in regions of star formation, as well as in the terminal stages of stellar evolution where they are found in most planetary and proto planetary nebulae. However, mundane, almost spherically symmetric winds emanate from many young star forming regions, as well as from massive main sequence stars, from the Sun, and from the AGB stars, which are commonly seen as large amplitude Mira variables. Yet our current understanding of the causality behind all of these mass-flows is slight.

Mira variables lose mass to interstellar space in low velocity winds. In most stars these are driven by radiation pressure acting on dust grains which condense within their flows at a few stellar radii (Gilman 1972; Kwok 1975; Jura 1984). Direct observational evidence implicating dust in this process includes (i) a correlation of mass-loss rate, $\dot{m}$, with UV opacity in carbon-rich circumstellar shells (CSs) (Knapp 1986); (ii) a strong increase in the expansion velocities, $v_{exp}$, of the winds from low mass Miras as their CSs become opaque, which facilitates the deposition of photon momentum (Lewis *et al* 1990); (iii) differences in the expansion velocities of carbon and oxygen rich stars that seem to depend only on envelope flavor (Lewis 1991); (iv) a difference in expansion velocities between OH/IR stars in the Galaxy and the LMC that are a result of their differing metallicities (Wood *et al* 1992). The dense winds from OH/IR stars of luminosity $L$ even carry a momentum flux, $\dot{m} v_{exp} = \tau (L/c)$ with $\tau >> 1$: this has been shown by Netzer & Elitzur (1993), by Ivezevik & Elitzur (1995), and independently by Habing, Tignon & Tielens (1994) to be explicable as a result of the multiple scattering of the stellar photon flux by dust particles. But 50 to 95% or so of the momentum flux, $\dot{m} v_{esc}$, needed to lift material from a Mira photosphere to infinity is consumed over the radial range below that at which dust forms, where the stellar atmosphere is essentially transparent.  Indeed some AGB stars, such as $\chi$ Cyg, have almost no dust, though they have winds.  Consequently most of the momentum flux needed to support AGB mass-loss must come from within the star itself, and so has no direct dependence on the existence of dust.

The detailed mechanisms for coupling momentum into a red giant's mass-loss stages remain uncertain (Hearn 1990), despite progress with the numerical simulation of mass-loss driven by shockwaves in distended stellar atmospheres (Bowen 1988).  Bowen's approach succeeds in pushing mass from the photosphere out to the cooler radii at which dust condenses: radiation pressure on the dust then sweeps the material away to infinity.  But Bowen's model assumes the existence of a periodic piston below the base of the atmosphere to inject the requisite momentum flux: the existence of a flux-source within the star is a given of his models.  Convincing *ab initio* calculations for the complete structure of AGB stars are still to come.  In consequence many simple questions about Miras remain unanswered, such as what triggers their characteristic pulsation (Wood 1990).  More fundamentally there is still confusion about exactly why a red giant star is a giant (Applegate 1988, Whitworth 1989).  However, none of the past work addresses the question of the physical origin of the momentum-flux carried in the wind from the photosphere, which can be likened to (and may be identical to) the absence of an accepted physical causality for the momentum flux carried by the solar wind (e.g. Holzer 1976).

In seeking an explanation for this flux I am drawn to the correlation between the occurrence of active phases of gravitational contraction in stars and the incidence of significant winds.  Spherically-symmetric, 5-30 km s$^{-1}$ winds occur during an early stage of protostellar evolution as well as the later stages of red giant evolution.  This focus on contraction is relevant to AGB stars, since they occupy the stellar evolution niche arising when steady contraction of the electron-degenerate core occurs, as helium ashes from shell-hydrogen burning make it ever more massive. Degenerate cores must contract under the addition of mass to maintain stability (Chandrasekhar 1939). Slow, spherically symmetric AGB winds may therefore be the reaction product of core contraction, whereon they would be the three dimensional analog of a rocket's exhaust. Similar winds occur during one stage of star formation (e.g. Shu & Lada 1990), while a protostar is separating itself from its molecular shroud. At a more advanced stage of the process proto-stellar winds usually seem to develop into bipolar flows, and the resulting T Tauri stars often have an annular disk of rotating matter. Bipolar flows occur in the last stages of AGB mass-loss too, and characterize most if not all proto planetary nebulae (Hrivnak 1996). Thus spherically symmetric winds seem to be a feature of active gravitational contraction when *angular momentum constraints are **not** important.*

The simplest systems with significant mass loss are those with little complexity added by rotation or magnetic fields, such as the AGB stars. We therefore explore the basic hypothesis here that the momentum flux required to drive a spherically symmetric AGB wind can be attributed to the excess to-virial-stability momentum flux generated by gravitational contraction after it has been transmitted through a star's center to become an outwardly directed flux. A similar inversion in direction occurs with standing waves in a spherical cavity (Landau & Lifshitz 1979). The central question posed by this paper concerns the proper treatment to accord this momentum flux, which is usually assumed to be cancelled by symmetry, without considering how that is achieved in detail.

Section 2 exhibits the present theory in outline form by calculating the momentum flux generated in the quasi-static contraction of a homogeneous sphere. A general formula for the flux from a contracting polytrope under the addition of mass is derived in section 3, and applied to AGB stars in section 4. A simplified theory is applied to the Sun in section 5 to explore the relationship of this approach to the momentum flux in the solar wind.

## 2. the secular contraction of a homogeneous sphere

The electron-degenerate core of an AGB star secularly increases in mass as helium "ashes" from shell hydrogen burning are added (e.g. Iben 1987): the core contracts to maintain virial equilibrium. Our objective is to quantify the resulting inwardly directed momentum flux for comparison with what is needed to sustain mass-loss. Both for simplicity, and to exhibit the essential features of the scenario quickly, our program is carried through in this section by treating the core as a homogeneous sphere.

Consider first a spherical shell of radius $r$ and mass $dm = 4\pi r^2 \rho\, dr$, with every particle in circular orbit about a central mass, $M(r)$, so no particle has a radially directed velocity component. The change in the potential energy of the shell, $\Delta\Omega$, as $r \to r - \Delta r$ in an infinitesimal contraction is

$$\Delta\Omega = GM\left(\frac{1}{r-\Delta r} - \frac{1}{r}\right)dm = dm\, v_{rot}^2(r)\left(\frac{\Delta r}{r-\Delta r}\right),$$

so its kinetic energy immediately after a radial contraction $\Delta r$ is to first order

$$T_{shell} = \tfrac{1}{2}dm\, v_{rot}^2(r) + dm\, v_{rot}^2(r)\left(\frac{\Delta r}{r-\Delta r}\right) = \tfrac{1}{2}dm\, v_{rot}^2(r)\left[1 + 2\Delta r/r\right].$$

The radially directed change in the velocity of every constituent particle is therefore

$$\Delta v = v_{rot}(r)\left[1 + 2\Delta r/r\right]^{1/2} - v_{rot}(r) \approx v_{rot}(r)\,\Delta r/r = \beta\, v_{rot}(r),$$

where $\beta$ is the fractional change in its radius. The resulting **radial** momentum flow has a magnitude $\Delta P = dm\, \Delta v$ (just as that resulting from the free-fall of a shell from infinity is $\Delta P(dm, r \leftarrow \infty) = dm\, v_{esc}$), since a net inward movement of mass constitutes a local momentum flow, despite the fact that spherically symmetric contraction does not change the vectorial sum of momentum in the system. This use of $\Delta P$ emulates the ubiquitous $L/c$ formulation (Salpeter 1974; Knapp *et al* 1982) for quantifying the total momentum flux carried by a star's luminosity in every radial direction.

We now extend this analysis to the simultaneous contraction of every shell of a stellar core in hydrostatic equilibrium. The resulting inwardly directed momentum change of each constituent shell is part of a coherent **radially inward momentum flow**. This cannot be cancelled by interactions within a mass distribution in hydrostatic equilibrium, where every other momentum component is by definition balanced, as Parker (1958) long ago emphasized. And inwardly directed flows have no possibility of being exhausted by work against gravity. The momentum flow must therefore reach the center.

The inwardly directed momentum increment generated by a proportional contraction of every radius of a homogeneous sphere by $\beta$ is thus

$$\Delta P = \int_0^{R_s} \beta v_{rot}(r) 4\pi r^2 \rho\, dr = \beta \int_0^{R_s} \left(\frac{4\pi r^3 \rho G}{3r}\right)^{1/2} 4\pi r^2 \rho\, dr = \frac{3\sqrt{2}}{8} \beta M v_{esc} \qquad (1).$$

Here $R_S$ is the surface radius, $M$ the total mass, and $v_{esc} = \sqrt{2GM/R_S}$ the escape velocity. When contraction is part of a continuous, quasistatic process, differentiation of $\Delta P$ with respect to time gives the momentum flux

$$\frac{dP}{dt} = \frac{3\sqrt{2}}{8} \frac{M}{R_S} v_{esc} \frac{dR_S}{dt} \qquad (2)$$

through the center engendered by a contraction rate $dR_S/dt$.

To progress further we need an expression for $dR_S/dt$. Our AGB core is represented at this point by a homogeneous sphere with the same surface radius as the core, whose luminosity, $L_*$, is related to its mass, $M_c$, in solar units by the relation

$$L_*/L_o = 52000\left(M_c/M_o - 0.456\right) \qquad (3)$$

deriving from stellar structure calculations (Boothroyd & Sackman 1992). This permits $\dot{M}_c$ to be obtained from $L_*$ by dividing it by the energy released per gm of hydrogen converted to helium, $E_H$, which after allowing for the mass fraction in hydrogen, $X$, gives

$$dM_c/dt = L_*/(X E_H) \qquad (4)$$

(Kwok 1987). The core contraction rate consequent on the mass increment rate $\dot{M}_c$ is got by differentiating the relation between $R_S$ & $M_c$ given in Hansen & Kawaler (1994; p127) as a fit to the results of exact calculations for degenerate white dwarfs

$$R_S/R_1 = 2.02\left[1-\left(M_c/M_\infty\right)^{4/3}\right]^{1/2}\left(M_c/M_\infty\right)^{-1/3} \qquad (5),$$

where $R_1 = 1.117*10^{-2}\,\mu_e^{-1}\,R_o$, $M_\infty = 1.456\,(2/\mu_e)^2\,M_o$, and $\mu_e$ the mean molecular weight per electron, so

$$\frac{dR_S}{dt} = -\frac{2.02\,R_1}{3\,M_\infty}\frac{dM_c}{dt}\left[\frac{1+\left(M_c/M_\infty\right)^{-4/3}}{\left\{1-\left(M_c/M_\infty\right)^{4/3}\right\}^{1/2}}\right] \qquad (6).$$

The minus sign signifies that the polytrope contracts on the addition of mass. Substituting (4) & (6) into equation 2 gives the desired expression

$$\frac{dP}{dt} = -0.2453 \left(\frac{R_1}{R_S}\right)\left(\frac{M_c}{M_o}\right) \left[\frac{1+(M_c/M_\infty)^{-4/3}}{\{1-(M_c/M_\infty)^{4/3}\}^{1/2}}\right] v_{esc} \left(\frac{L_*}{X E_H}\right) \quad (7).$$

The momentum flux needed for lifting mass from the photosphere to the radius at which dust condenses is $\sim \dot{m} v_{esc}$. For the canonical Mira of Humphreys *et al.* (1996), with an effective surface temperature $T_{eff}$ = 3000 K, period of 332 days, $R_S$ = 244 $R_o$, expansion velocity $v_{exp} \approx$ 10 km s$^{-1}$, and $\dot{m}$ = 1.8 10$^{-7}$ $M_o$ yr$^{-1}$, $L_*$ = 4340 $L_o$, which from equation (3) implies an $M_c$ = 0.539 $M_o$. This has a photospheric $v_{esc}$ = 39.5 km s$^{-1}$, so for $M$ = 1.0 $M_o$ (core + envelope)

$$\dot{m} v_{esc} = 0.081 \left(L_* / c\right) \quad (8).$$

This should be compared with the contraction flux calculated from equations 4 & 7 with $X$ = 0.7 and $E_H$ = 6 10$^{18}$ ergs gm$^{-1}$, of

$$\frac{dP}{dt} = 0.850 \left(L_* / c\right) \quad (9).$$

Clearly the momentum flux engendered by core contraction is in this case an order of magnitude larger than that directly needed to sustain the modest mass-loss rate in (8). In the absence of an efficient central momentum cancellation mechanism, this flux passes through the center to become in turn an outwardly directed flux that is transmitted from the core to the stellar envelope, to promote and sustain its characteristic pulsation mode. This in turn leads to a shock distended stellar atmosphere, and thus to photospheric mass loss (Bowen 1988).

## 3. Contraction Momentum from Polytropes

White dwarfs have a polytropic mass distribution (Chandrasekhar 1939). Our next objective is accordingly to quantify the momentum flux generated by a polytrope under changes to its radius, $R_S$, and mass, $M$. We begin with the gravitational potential energy, $\Omega$, of a stable polytrope of index n (Chandrasekhar, p101), which is related to its total kinetic (or thermal) energy content, $T$, through the Virial Theorem such that $2T + \Omega = 0$,

$$\Omega = -\left(\frac{3}{5-n}\right)\frac{GM^2}{R_S},$$

so
$$T = -\frac{1}{2}\Omega = \frac{1}{2}\left(\frac{3}{5-n}\right)\frac{GM^2}{R_S} = \frac{1}{2}M\lambda_n^2 v_{esc}^2 = \frac{1}{2}M v_{rms}^2,$$

with $\lambda_n = \left[\frac{1}{2}\left(\frac{3}{5-n}\right)\right]^{1/2}$. The rms velocity, $v_{rms}$, of particles (for the moment assumed to all be of one kind) is related to their mean velocity, $\bar{v}$, in a polytrope by

$$\bar{v} = \lambda v_{rms} = \lambda \lambda_n v_{esc} \qquad (10),$$

where $\lambda = 0.921$ when they are from a Maxwell-Boltzmann distribution. Let us define $P$, the total "directionless" or scalar momentum content of a polytrope in virial equilibrium, as

$$P = M\bar{v} = M\lambda \lambda_n v_{esc} \qquad (11),$$

which relates to the usual expression between momentum and kinetic energy via

$$\frac{P^2}{2M} = \lambda^2 T \qquad (12).$$

The value of $\lambda$ depends on the motion of constituent matter. For an n = 0 polytrope (the homogeneous sphere) with every constituent particle in circular orbit about the center, following the development of (1) above, $\bar{v}$ is

$$\bar{v} = \frac{1}{M} \int_0^{R_S} v_{rot}(r) \, 4\pi r^2 \rho \, dr = \frac{1}{M} \int_0^{R_S} \left(\frac{GM(r)}{r}\right)^{1/2} 4\pi r^2 \rho \, dr = \frac{3\sqrt{2}}{8} v_{esc} \quad (13),$$

and since (10) defines $\bar{v} = \lambda \lambda_0 v_{esc}$, $\lambda$ is then $\frac{3\sqrt{2}}{8}\left(\frac{1}{\lambda_0}\right) = 0.559$.

The homogeneous sphere has utility in treating the momentum content of polytropes, as it is easily integrated. The results are then generalizable to an arbitrary polytrope of index n, since this can always be mapped to the equivalent homogeneous sphere with the same total binding energy and mass, by evaluating its surface radius, $R_{S,o}$, via

$$T = -\frac{\Omega}{2} = \frac{1}{2}\left(\frac{3}{5-n}\right)\frac{GM^2}{R_{S,n}} = \frac{1}{2}\left(\frac{3}{5}\right)\frac{GM^2}{R_{S,o}},$$

so
$$R_{S,o} = \frac{5-n}{5} R_{S,n} = \left(\frac{\lambda_o}{\lambda_n}\right)^2 R_{S,n} \quad (14).$$

It follows that $\lambda_n v_{esc,n} = \lambda_o v_{esc,o}$, whereon the momentum content of the polytrope, $P_n$, is

$$P_n = M\bar{v}_n = M\lambda\lambda_n v_{esc,n} = M\lambda\lambda_o v_{esc,o} = M\bar{v}_o = P_o \quad (15).$$

This gives the exceedingly useful result that the $P$ of a polytrope of index n is readily derived from that of its equivalent homogeneous sphere.

The second use of $P$ comes from its differentiability: that with respect to radius gives

$$dP = \lambda \lambda_0 M \frac{dv_{esc}}{dR_S} dR_S = -\frac{1}{2} \frac{3\sqrt{2}}{8} \frac{M}{R_S} v_{esc} dR_S = -\frac{1}{2} \frac{P \, dR_S}{R_S} = -\frac{1}{2} \beta P \qquad (16),$$

while a similar differentiation of the $T$ of a poytrope gives

$$dT = d\left(\frac{1}{2} M \lambda_n^2 v_{esc}^2\right) = -T \frac{dR_S}{R_S} = -\beta T \qquad (17).$$

Both $dP$ and $dT$ are positive when account is taken of the **decrease** in radius attendant on contraction. So (16) & (17) show that $dP = 0.5 (P/T) dT$ is proportional to the change in the kinetic energy of the polytrope. It should be noted that while (16) gives the change in the momentum content of a polytrope in virial **equilibrium**, $dP = -0.5 \beta P$, (1) gives the change $dP = -\beta P$ resulting from an instantaneous contraction before any contraction energy is exported. The factor of two difference in coefficients

$$\frac{0.5 \beta P}{\beta P} = \frac{1}{2} = \frac{\Delta T}{\Delta \Omega}$$

is simply the ratio between the change in retained kinetic energy of a polytrope in equilibrium and the change in its potential energy under contraction, as half of the contraction energy in non-relativistic regimes must be exported to maintain virial equilibrium. In consequence a summation over the **inwardly directed momentum** generated by quasistatic contraction through $-\Delta R_S$ has a magnitude of *twice the change in the equilibrium value of P for the polytrope:* this is exactly quantified by the differentiation of $P$ with respect to $R_S$

The rate of change in the $P$ of a polytrope under changes to both its mass and radius is obtained by differentiating (11)

$$\frac{dP}{dt} = \frac{\partial P}{\partial M}\frac{dM}{dt} + \frac{\partial P}{\partial R_S}\frac{dR_S}{dt} = \frac{3\bar{v}}{2}\frac{dM}{dt} - \frac{\bar{v}}{2}\frac{M}{R_S}\frac{dR_S}{dt} \qquad (18).$$

The sign of the second term in (18) is cancelled during contraction by that of $dR_S/dt$, so the terms add and will be seen below to be of comparable size. But the $\Delta P$ generated by the addition of $\Delta M$ is at most $\Delta P = 2(\Delta M/M)P$ (Appendix A, equation A2) when $\Delta M$ is brought from infinity, whereas (18) shows the change to the $P$ of the polytrope alone under the addition of mass at constant $R_S$ is $(3/2)(\Delta M/M)P$: a doubling of the first term in (18) does not in this case provide its contribution to the total momentum generated by contraction. Thus the mode of mass addition needs more scrutiny.

Mass is added to an AGB core from its periphery as it undergoes quasistatic contraction, so it is appropriate to consider the $\Delta M$ mass addition as coming from a thin shell at a radius $(R_S + \Delta r)$ with a **prior** kinetic energy $\Delta T = 0.5\,\Delta M\, v_{rot}^2(R_S + \Delta r)$ and a prior $\Delta P(\Delta M) = \Delta M\, v_{rot}(R_S + \Delta r)$. This can be rewritten in the case of an n=0 polytrope as

$$\Delta P(\Delta M) = \frac{\sqrt{2}}{2}\Delta M \left(\frac{2GM(r)}{R_S + \Delta r}\right)^{1/2} = \frac{4}{3}\left(\frac{\Delta M}{M}\right)P_o \qquad (19),$$

where $P_o = \frac{3\sqrt{2}}{8} M v_{esc}$ from (11) & (13). The difference

$$\delta P = \frac{3}{2}\left(\frac{\Delta M}{M}\right)P - \frac{4}{3}\left(\frac{\Delta M}{M}\right)P = \frac{1}{6}\left(\frac{\Delta M}{M}\right)P \qquad (20).$$

between the net change $\Delta P = 1.5(\Delta M/M)P$ of the mass-augmented polytrope in virial equilibrium at constant $R_S$ after the addition of $\Delta M$ from (18), and the $\Delta P$ of (19) which is the momentum content of $\Delta M$ while it is still at $(R_S + \Delta r)$, is the increase in the polytrope's $P$ due to homologous contraction as $M(R_S) \to M(R_S - \Delta r)$ and $\Delta M(R_S + \Delta r) \to \Delta M(R_S)$. Since (20) originates solely from contraction, it too represents a radially directed momentum flow. It follows from (18) & (20) that the scalar sum of "freshly generated" momentum from a polytrope in **virial equilibrium**, consequent on changes to its mass and radius, is

$$\Delta P = \frac{1}{6} \bar{v} \Delta M - \frac{1}{2} \bar{v} \left( \frac{M}{R_S} \right) \Delta r \qquad (21).$$

Differentiating (21) to cast it as a flux, and doubling its coefficients to get the inwardly directed momentum flux, gives the basic momentum flux relation for any polytrope

$$\frac{dP}{dt} = \frac{1}{3} \bar{v} \frac{dM}{dt} - \bar{v} \left( \frac{M}{R_S} \right) \frac{dR_S}{dt} \qquad (22).$$

Equations 4 & 6 are then used in conjunction with (22) to predict the momentum flux arising from the quasistatic contraction of an AGB core.

## 4. Application to AGB stars

This section deals first with the momentum flux from the core of an AGB star, then with what is needed for mass-loss from its photosphere, before finishing with the results accruing from applying this approach to understanding the winds of AGB stars.

**4.1: on the momentum flux from core contraction**

The **whole** of a degenerate core contracts in reaching virial equilibrium on the addition of mass, and half of the resulting contraction energy is exported (less is exported from highly relativistic cores). And though the instantaneous scalar momentum change $\Delta P \propto \Delta \Omega$, the loss of ≈50% of the contraction energy does not of itself imply an exactly proportionate export or cancellation of 50% of $\Delta P$, since this outcome is mediated by (i) equipartition of energy between particle species, and (ii) the core is supported against collapse by electron degeneracy rather than thermal pressure.  We look at these factors next.

(i) *equipartition of energy*

Suppose for simplicity a non-degenerate core composed entirely of carbon-12 nuclei, so there are 6 electrons for every nucleus.  Under equipartition of energy the kinetic energy of all nuclei in a polytrope in thermal equilibrium is

$$T(C_{12}) = \frac{T}{7} = \frac{1}{7} \frac{M}{2} \lambda_n^2 v_{esc}^2 = \frac{1}{2} \left( \frac{m_{12}}{m_{12} + 6 m_e} \right) M v_{rms}^2 ,$$

whereon $$v_{rms,nuc} = \left( \frac{1}{7} \frac{m_{12} + 6 m_e}{m_{12}} \right)^{1/2} \lambda_n v_{esc} ,$$

so for the nuclei alone $$P_{nuc} = \left( \frac{m_{12}}{m_{12} + 6 m_e} \right) M \bar{v}_{nuc} = \left( \frac{1}{7} \frac{m_{12}}{m_{12} + 6 m_e} \right)^{1/2} P \quad (23),$$

while the contribution from electrons is $$P_e = \left( \frac{6}{7} \frac{6 m_e}{m_{12} + 6 m_e} \right)^{1/2} P .$$

The ratio of the scalar momentum content carried by electrons to that by nuclei is $6 (m_e / m_{12})^{1/2}$ = 0.04, and the ratio $P_{nuc} / P$ = √1/7 = 0.38, or $(\mu / \mu_I)^{1/2}$ whenever the number of electrons per nucleus is ≈$\mu_I$/2, where $\mu$ and $\mu_I$ are respectively the mean molecular weight and the mean molecular weight of ions.  Clearly the equipartition of energy between particle species results in the scalar momentum content of a polytrope in equilibrium being much smaller than if it were composed of just one species.

*(ii) the scalar momentum content of nuclei in a degenerate core*

Since core nuclei are supported against gravity by electron degeneracy rather than thermal pressure, their kinetic temperature is markedly lower than that needed for virial equilibrium under thermal pressure. And their temperature, $T_{nuc}$, is necessarily less than the $T_{He} \approx 10^8$ K that triggers conversion of $He^4$ to $C^{12}$ during shell-hydrogen burning phases of an AGB star. The cores are therefore nearly isothermal. The mean kinetic energy of a carbon-12 nucleus is $(1/2) m_{12} v^2_{rms,nuc} = (3/2) k T_{nuc}$, so the scalar momentum content, $P_{nuc} = N m_{12} \bar{v}_{nuc} = N m_{12} \lambda v_{rms,nuc}$ of an entire core is

$$P_{nuc} = (M N_A / \mu_I) \lambda \sqrt{3 k T_{nuc} m_{12}} \qquad (24),$$

where $k$ is Boltzmann's constant, $N_A$ Avogadro's number, and $N = M N_A / \mu_I$ the total number of ions in the core. Since nuclei have a Maxwell-Boltzmann velocity distribution, $\lambda$ is in this instance 0.921.

*(iii) the scalar momentum content of the electrons*

We derive an approximate value for the scalar momentum content carried by the electrons of a degenerate core, when this is treated as a homogeneous sphere next: this follows the treatment Hansen & Kawaler (1994, p124ff) accord the Chandrasekhar mass limit. The electron number density, $n_e$, of a completely degenerate, zero-temperature, non-relativistic sphere whose Fermi-momentum $p_F = x_F (m_e c)$, is

$$n_e = \frac{8\pi}{h^3} \int_o^{p_F} p^2 \, dp = \frac{8\pi}{3 h^3} p_F^3 = \frac{8\pi}{3} \left(\frac{h}{m_e c}\right)^{-3} x_F^3 \qquad (25),$$

where $h$ is Planck's constant. But $n_e = \rho N_A / \mu_e$, so for a homogeneous sphere $x_F$ is

$$x_F = \left[\frac{9 M N_A}{32 \pi^2 \mu_e}\right]^{1/3} \frac{1}{R_S}\left(\frac{h}{m_e c}\right) \qquad (26).$$

The scalar momentum content of the electron distribution, after substituting (26), is

$$P_e = \frac{4}{3}\pi R_S^3 \int_o^{P_F} \frac{8\pi}{h^3} p\, p^2\, dp = \frac{8\pi^2}{3}\left(\frac{h}{R_S}\right)\left\{\frac{9 M N_A}{32 \pi^2 \mu_e}\right\}^{4/3} \qquad (27).$$

This provides an upper limit to $P_e$, since all other polytropes have a larger central density, which increases the mean electron energy in the center and reduces their contribution to $P_e$. When (27) is evaluated for our prototype Mira and $P_{nuc}$ is calculated on the same basis from (24), $P_e / P_{nuc} = 0.08$. Though most of $P$ resides in the ions, this changes as the core mass increases, as a result of the concomitant decrease of $R_S$ on (27).

The scalar momentum content of a body in virial equilibrium is the sum of contributions from its ions and electrons, $P_{core} = P_{nuc} + P_e$. This is smaller than the $P$ defined by (11) when the existence of more than one kind of constituent particle is ignored. For our prototypical Mira core of §2, as it contracts through $-\Delta R$ under the addition of mass

$$\Delta P_{core} = \left\{(\Delta P_{nuc} / \Delta P) + (\Delta P_e / \Delta P)\right\} \Delta P \approx \{0.1012 + 0.0081\} \Delta P \approx 0.11\, \Delta P,$$

where $\Delta P$ is the total change in $P$ engendered by core contraction from (22). It is clear from this calculation that ≈90% of the momentum generated in quasistatic contraction is either exported from or cancelled in the core in maintaining equilibrium.

The momentum flux equation (22), on substitution of (4) & (6) becomes

$$\frac{dP}{dt} = \frac{\bar{v}}{3}\left(1 + \frac{2.02\, R_1}{R_{S,c}}\frac{M_c}{M_\infty}\left[\frac{1+(M_c/M_\infty)^{-4/3}}{\{1-(M_c/M_\infty)^{4/3}\}^{1/2}}\right]\right)\left(\frac{L_*}{X\, E_H}\right) \quad (28),$$

where $M_c$ is the mass and $R_{S,c}$ the surface radius of the core. This is evaluated with $\bar{v} = \lambda\, \lambda_n\, v_{esc,n} = (3\sqrt{2}/8)\sqrt{5/(5-n)}\, v_{esc,n}$, in accord with (10) & (15). It follows from the discussion above that a fraction $\varepsilon = 1 - (P_{nuc} + P_e)/(2P)$ of this radially directed momentum flux is surplus to the stability of the core.

Table I lists some representative values of the momentum flux generated by different core masses when (28) is evaluated with $X = 0.7$, n = 3/2, and $T_{nuc} = 10^8$ K, together with (3). The flux increases much more quickly than the luminosity with $M_c$, as is evident in its order of magnitude increase down the table when expressed in units of $L_*/c$ of the stellar luminosity divided by the velocity of light, which already discounts the associated increase in luminosity. This portion of the flux increase is due in part to a modest increase in the fraction available for export as $\varepsilon$ increases with $M_c$. But most of it is due to the accompanying decrease in $R_{S,c}$, which causes a rapid increase in the second (radial contraction) term in the main bracket of (28), whose relative value increases by a factor of 5 down the table. In real cores this effect is enhanced further by n increasing with $M_c$.

### 4.2: momentum flux requirements of a slow wind

The second part of the stellar wind problem is to determine what momentum flux, $\dot{P}$, is implied by the occurrence of mass-loss from a photosphere in the *ideal* circumstance where none is cancelled by shocks in the distended atmosphere: more is of course needed to compensate for these losses. The **smallest** momentum input enabling mass loss from a photosphere corresponds to mass leaving at a velocity sufficient to reach the radius of dust formation, $r_d$. When it is further assumed that $r_d = \infty$, we have the impulsive limit $\dot{P} = \dot{m}\, v_{esc,p}$, where $v_{esc,p}$ is the escape velocity from the photosphere. But Bowen's mass-loss models have mass leaving the photosphere at a smaller radial velocity in a more massive mass flux, which requires a larger $\dot{P}$. Quasistatic contraction has the opposite bias to that of the impulsive limit, as it maximizes the quantity of radially directed momentum generated during contraction. We therefore use its formalism next to place an upper limit on the $\dot{P}$ required to sustain mass-loss from a photosphere.

During quasistatic contraction of a polytrope from $r_d$ to $R_S$, the change to its $P$ is

$$P(R_S) - P(r_d) = \frac{3\sqrt{2}}{8}\left(\frac{\lambda_n}{\lambda_o}\right) M \sqrt{2GM}\left(\frac{1}{\sqrt{R_S}} - \frac{1}{\sqrt{r_d}}\right) \quad (29)$$

from (11). Similarly, when the polytrope has instead a mass $M - dM$

$$P(R_S) - P(r_d) = \frac{3\sqrt{2}}{8}\left(\frac{\lambda_n}{\lambda_o}\right)(M - dM)\sqrt{2G(M-dM)}\left(\frac{1}{\sqrt{R_S}} - \frac{1}{\sqrt{r_d}}\right) \quad (30),$$

so the difference between (29) & (30), retaining first order terms, is

$$\Delta P(\Delta M) = \frac{3}{2}\frac{\Delta M}{M}\left[P(R_S) - P(r_d)\right] \quad (31).$$

Finally by adding to (31) the term from (20), which allows for the momentum-debt of subtracting mass from a polytrope while $R_S = r_d$, we get the quasistatic change to the $P$ of a polytrope when it expands from $R_S$ to $r_d$ with mass $M$, loses $dM$, and then contracts to $R_S$. These operations on the equilibrium values of $P$ for a polytrope represent just **half** of the total scalar momentum change implied by this quasistatic cycle, whose net result is the loss of $dM$ at zero velocity from $r_d$. The momentum flux required to quasistatically power a wind (the time-reversed inverse of contraction) is therefore

$$\Delta P(\Delta M) = \frac{1}{3}\frac{\Delta M}{M}\left[9\,P(R_S) - 8\,P(r_d)\right] \quad (32).$$

In the interesting limit of $r_d \to \infty$, (32) expressed as a flux reduces to

$$\dot{P} = 3\frac{3\sqrt{2}}{8}\left(\frac{\lambda_n}{\lambda_o}\right)\dot{m}\,v_{esc,p} \quad (33),$$

which gives $\dot{P} = 1.9016\,\dot{m}\,v_{esc,p}$ for an n = 3/2 polytrope.

The two momentum flux bounds on mass-loss from the photosphere as $r_d \to \infty$ are

$$\dot{m}\,v_{esc,p} < \dot{P} < 1.9\,\dot{m}\,v_{esc,p} \quad (34).$$

These are both made smaller by taking detailed note of $r_d$, as mass is then lifted through a smaller height against gravity before radiation pressure on dust takes over, whereon

$$\dot{m}\,v_{esc,p}\left(1 - \sqrt{\frac{R_S}{r_d}}\right) < \dot{P} < \sqrt{2}\,\dot{m}\left(\frac{\lambda_n}{\lambda_o}\right)v_{esc,p}\left(\frac{9}{8} - \sqrt{\frac{R_S}{r_d}}\right) \quad (35).$$

When a model dependent $r_d$ is related to a usual dust condensation temperature ≈1000 K in an oxygen-rich shell (e.g. Humphreys *et al* 1996) via $r_d = R_S(T_{eff}/1000)^{5/2}$, and $T_{eff} \approx 3000$ K, $r_d$ is 15.6 $R_S$. This gives our 1 $M_o$ Mira of §2 the momentum flux bounds

$$0.747\,\dot{m}\,v_{esc,p} < \dot{P} < 1.420\,\dot{m}\,v_{esc,p} \quad (36).$$

The $\dot{P}$ needed for lifting mass from $R_S$ to $r_d$ is set by these radii. These depend on $T_{eff}$,

whose value is correlated with both $\dot{m}$ and progenitor mass. Most Miras have almost translucent shells with $T_{eff} \approx 3000$ K, while heavily shrouded objects have much lower temperatures, though the actual $T_{eff}$ are moot. Le Sidaner & Le Bertre (1996) assign $T_{eff}$ of 1800 - 2200 K to OH/IR stars, whereas a very similar shell-modelling study by Lepine, Ortiz & Epchtein (1995) assigns many of them a $T_{eff} \approx 1250$ K. WX Psc is an intermediate case. It is a heavily shell-reddened star with an M9.5-10 spectral type (Lockwood 1985), and so perhaps may be accorded a $T_{eff} \sim 1600$ K (Dyck *et al* 1974). We expect the reddest OH/IR stars to have the lowest $T_{eff}$, which is supported by the correlation between spectral type and expansion velocity (Lewis 1991). In practise the stars with the largest $\dot{m}$ have the coldest photospheres, which means they raise their mass flux through a proportionately smaller radial range to reach $r_d$. These objects therefore have the innate capacity to support the largest $\dot{m}$ when mass-loss is a momentum bounded process. Table 2 illustrates these trends for a 1 $M_O$ star. Its last column shows that a 3.4 times larger momentum flux is needed to drive the same $\dot{m}$ from a star with $T_{eff} = 3000$ K as from one with $T_{eff} = 1500$ K.

### 4.3: results for AGB stars

We saw in §2 that contraction readily provides for the needs of an $\dot{m} \approx 1.8 \ 10^{-7} \ M_O \ yr^{-1}$ loss from our prototype Mira with $T_{eff} \approx 3000$ K. But an important fraction of any momentum flux is cancelled in photospheric shocks, and some may also be cancelled at the center, so a considerable extra capacity for generating a flux, over and above the direct needs of the stellar wind, is a mandatory feature of any theory. We work hereafter from the basis that the minimal flux is specified by the quasistatic upper bound of (35). The most demanding oxygen-rich AGB stars to be accounted for have an $\dot{m} \sim 10^{-4} \ M_O \ yr^{-1}$ (Knapp & Morris 1985). For an assumed $T_{eff} = 1500$ K, the quasistatic bound for sustaining this wind from a 1 $M_O$ star with $L = 30,000 \ L_O$ is 1.76 $L_* / c$; if $T_{eff} = 1750$ K ,

the bound is 2.47 $L_*/c$. These bounds are both suitably smaller than the $\dot{P}$ = 5.51 $L_*/c$ listed in column 5 of Table 1 as being generated by core contraction at this luminosity. Since the AGB stars with the largest $\dot{m}$ may have $L \approx 50{,}000\ L_0$, and so may generate a commensurately larger $\dot{P}$, we conclude that core contraction is easily able to provide for the momentum flux needs of every AGB wind.

An immediate result of attributing the AGB wind's momentum flux to gravitational contraction is its consequent sensitivity to $M_c$, and thus to $L$. AGB stars with small progenitor masses, and so relatively small luminosities, absolutely can not achieve the largest observed mass-loss rates. This has already been forced on our notice by observations of circum-stellar masers that show a marked decrease in the galactic latitude distribution of OH/IR stars with increasing shell opacity, and hence $\dot{m}$, when these parameters are assessed from their IR colors (Lewis 1987). An update of this evidence is presented in Figure 1 for the IRAS sources with detected OH masers: Figure 1 shows an increasingly marked paucity of high latitude objects when (25-12) μm > -0.4, as shells get increasingly redder and more opaque. Column 6 in Table 1 illustrates the same point numerically in documenting the rapid increase in the maximum $\dot{m}$ with $M_c$. This is obtained by equating the $\dot{P}$ calculated with (28) & (3) to three times the quasistatic shell bound computed for $T_{\mathit{eff}}$ = 1500 K: the factor of 3 here is equivalent to the assumption that just one third of the generated momentum flux survives cancellation in atmospheric shockwaves to support mass-loss at all progenitor masses. On this basis an $\dot{m} \approx 10^{-5}\ M_o$ yr$^{-1}$ implies an $L \approx 10{,}000\ L_o$ and $M_c \sim 0.65\ M_o$.

Some predictions flow from the applicability of our premise. Thus a *sine qua non* for the occurrence of mass-loss from an AGB star is that its core must be contracting. Yet during a thermal pulse the outer helium shell is stripped from a core as it burns, causing a net loss of core-mass. The core must then expand to maintain virial stability through the event, and so becomes a net consumer of momentum, which implies that $\dot{m} \approx 0$ during this phase. Indeed CO observations of U Ant, S Sct, & TT Cyg show substantial decreases in their mass-loss rates in the wake of a thermal pulse (Olafsson *et al* 1993).

Part of this is attributable to temporarily smaller core masses, but most is due to the drop in luminosity that follows on a thermal pulse (Wood & Zarro 1981), which implies a correspondingly lower rate of mass accretion. On the other hand post-AGB stars at the termination of mass-loss, following the dissipation of most of the stellar envelope, continue to burn hydrogen as rapidly as ever, and so continue to increase their core-masses. Mass-loss should therefore continue in some form as the most efficient mode for the dissipation of $\dot{P}$, even if the nature of the wind and its acceleration changes. "Fast winds" have long been observed to emanate from the central stars of planetary nebulae.

There are natural extensions of these ideas to stars with angular momentum (and/or strong magnetic fields), since the second *sine qua non* of the mechanism is point symmetry rather than spherical symmetry in accretion. A degenerate core spins faster as it grows in mass and contracts, so the angular momentum of material will at some stage affect the local accretion rate. This is necessarily slower at the equator than the poles, so an erstwhile spherical symmetry in the production of a momentum flux develops axial symmetry. This sculpts the wind, which refashions its environs. Material is then swept away from the poles more quickly than from the equator as there is a larger flux there, which promotes the probability of both equatorial disks and bipolar flows. These factors operate in every AGB star (and in every mass accreting object), not just those with stellar companions. The bipolar structure of most planetary nebulae (Zuckerman & Aller 1986) and proto planetary nebulae (Hrivnak 1996) is therefore an expected result of the applicability of this theory. It may also be applicable to star forming regions, since these mirror the same progression from spherically symmetric winds to bipolar flows.

## 5. The Solar Wind

Red supergiants have slow dusty winds rivaling those from AGB stars without having degenerate cores. By analogy they must also generate momentum fluxes, which in this case occur as much as a byproduct of stellar fusion as of contraction. Since the central zone of a star contracts to maintain thermal pressure as its particles are consumed in nuclear reactions, this effect may suffice to generate the $\dot{P}$ required to sustain mass loss from red supergiants, a supposition that needs testing with a stellar code. We use the idea here with a simple polytropic model of the Sun to see how it relates to the solar wind.

Let us model the Sun with an n = 3.25 polytrope. This maps into the homogeneous sphere with the same binding energy via (14), so we can predict a contraction rate by assuming it maintains a constant particle number density as fusion reduces its particle count. The assembly of each He$^4$ nucleus reduces this by 5, so $dN/dt = \frac{5}{4} N_A (L_o / E_H)$. The rate of volume change of the homogeneous sphere is then

$$\frac{d(\frac{4}{3} \pi R_S^3)}{dt} = \frac{5}{4} N_A \left(\frac{L_o}{E_H}\right) \frac{\mu}{\rho N_A},$$

where $\mu$ is the mean molecular weight and $\rho = M_o / \frac{4}{3} \pi R_S^3$, whereon

$$\frac{dR_S}{dt} = \frac{5}{12} \mu \left(\frac{R_S}{M_o}\right)\left(\frac{L_o}{E_H}\right) \quad (37).$$

This relation is unchanged by mapping between the sphere and polytrope. Substitution into (2) gives the momentum flux, $\dot{P}_o$, implied by homologous contraction of the solar polytrope, of which a fraction $\varepsilon$ is available for export following an argument like §4.1(i), so

$$\frac{dP}{dt} = \varepsilon \frac{5\sqrt{2}}{32} \mu \left(\frac{\lambda_{3.25}}{\lambda_o}\right) v_{esc,3.25} \left(\frac{L_o}{E_H}\right) \approx 0.177 \, v_{esc,3.25} \left(\frac{L_o}{E_H}\right) \quad (38).$$

The numerical value comes from using $\mu \approx 0.613$, and $\varepsilon = \{1 - (X/2 + Y/3 + Z/3) / 2\} \sim 0.774$, in which $X$, $Y$, & $Z$, the mass fractions of H, He and heavy elements, are respectively 0.7094, 0.271 & 0.0196. For $v_{esc,3.25}$ = 617 km s$^{-1}$, $\dot{P}_o$ is 7.0 10$^{21}$ gm cm s$^{-2}$ (i.e. 0.055 $L_* / c$). The $\dot{P}$ powering a quiescent solar wind at a velocity ≈400 km s$^{-1}$ and an $\dot{m} \approx 10^{-14}$ $M_o$ yr$^{-1}$ (Cassinelli & MacGregor 1986) is from (35) with n = 3/2

$$\dot{P}_{o,wind} = \sqrt{2}\dot{m}(\lambda_n/\lambda_o)[v_{esc,3.25} + 400] 9/8 = 1.219 \ 10^{20} \ gm \ cm \ s^{-2} \quad (39).$$

Thus $\dot{P}_o$ is ≈57 times larger than the quiescent solar wind needs.

The momentum flux required to sustain the totality of components subsumed into the solar wind is uncertain, though it is likely to be 2-3 times larger than (39). Thus apart from the flux originating from flares and other impulsive outbursts, Thomson *et al* (1995) discuss evidence linking an additional impulsive component of the wind to standing waves in the solar structure. These are omnipresent, encompass the whole surface, and may on their own more than double the particulate flux emerging from the Sun. Even so the $\dot{P}_o$ from (38) is still an order of magnitude larger than the whole of the solar wind needs.

The present calculation is simplistic since the radius of the Sun increases slowly as it evolves. This reflects the influence of other factors on its structure, including a very modest rise in central temperature, which decreases the available momentum flux and opposes the tendency for the whole mass distribution to contract. But the basic structure of the Sun is determined by hydrostatic equilibrium, when the tiny perturbation caused by fusion is ignored, so the contraction momentum flux must either be transmitted to the surface and/or used to expand some of the layers overlying the nuclear burning zone. The differential importance of these options needs elucidation from a stellar code.

It is instructive to compare $\dot{P}_o$ with the magnitude of the momentum cancellation rate inherent in stellar fusion. The cancellation rate from the dominant, non-resonant, p-p reaction can be approximated by assuming it occurs at the Gamow peak energy, $\varepsilon_o = 1.22 \left( Z_1^2 Z_2^2 \mu_r T_6^2 \right)^{1/3}$, where $\mu_r$ is the reduced mass of the interacting particles (1,2) in amu, $T_6$ in units $10^6$ K, and $\varepsilon_o$ in kev. The momentum cancelled per He atom is $\Delta P(\text{He}^4) \approx 2\sqrt{m_p \varepsilon_o}$ , which leads to a solar p-p cancellation rate when $T_6 = 15$ of

$$\frac{dP}{dt} = \frac{N_A}{2}\sqrt{m_p \varepsilon_o}\left(\frac{L_o}{E_H}\right) = 2.43\ 10^{22}\ gm\ cm\ s^{-2} \quad (40).$$

Since (40) is much larger than (38), there will normally be **NO** equality between the flux generated by contraction and that cancelled in fusion. In actuality momentum cancelled in fusion reactions is returned to the stellar structure as fusion energy is thermalised. Since fusion is much the most efficient stellar channel for cancelling momentum, this result suggests the contention that *the scalar momentum content of a stellar structure is statistically constant so long as its mass distribution and its ionised mass-fraction remain unchanged.*

## 6. discussion

Gravitational contraction always generates an inwardly directed momentum flux. The magnitude of this flux is quantified here for the quasistatic contraction of a polytrope, which is well suited to describing the secular contraction of an electron degenerate core under the addition of helium ashes: it is also applied to the contraction of the nuclear burning zones of the Sun as its particle number count changes with fusion. The momentum flux generated by gravitational contraction is, on the assumption that it is all transmitted through the center to become in turn an outwardly directed flux, more than sufficient to provide for the momentum requirements of slow, $\dot{m} \approx 10^{-4}$ $M_o$ yr$^{-1}$, AGB winds on the one hand, as well as those needed by the $\dot{m} \approx 10^{-14}$ $M_o$ yr$^{-1}$ solar wind on the other. A mass efflux is probably a general concomitant to gravitational contraction whenever a system has insufficient resources for cancelling its radially coherent momentum flux. An alternate resource is pulsation, which allows radially directed momentum to be thermalised in atmospheric shockwaves. The known correlation between pulsation and mass-loss is therefore explicable as a correlation between independent pointers to the existence of a significant, radially directed momentum flux.

The application of this causality to the specific case of AGB stars explains the extreme sensitivity of their maximum mass-loss rates to progenitor mass. Indeed the momentum flux from their degenerate cores has a magnitude 1 - 10 times that of the radiant flux in Table 1, which necessarily has implications for the structure of their envelopes. This is certainly one of the reasons Red Giants are giants, as their contracting cores are sources of a significant momentum flux that has to be absorbed and/or dissipated by the envelope. Since the essential feature underwriting the ability of the contraction mechanism to generate a momentum flux is point symmetry about the center, any important axially symmetric feature of the core, such as its rotation or magnetic field, can influence the local accretion-rate onto the core. This in turn impresses an axial signature on its contraction momentum flux, which will lead in turn to large scale structural features, such as bipolar flows and equatorial disks. The prevalence of bipolar flows in both proto planetary nebulae and planetary nebulae is thus an explicable consequence of the export of the contraction momentum flux from a rapidly spinning core.

These conclusions follow from the realization that there are several ways in which a spherically symmetric momentum flux can be treated, so that the overiding principle of conservation of momentum is satisfied. One mode is for it to be explicitly cancelled at the center of symmetry, such as happens with single bubble sonoluminesence. But no analogous mechanisms are available to stellar systems. A second mode is for it to be thermally cancelled near the center, which is difficult within the constraints of hydrostatic and thermal equilibrium obtaining in stellar structures. It is, however, reasonably easy to achieve some cancellation by disturbing the local equilibrium at the periphery of a star. This can be facilitated by pulsation, as happens with the ejection of energy into the solar wind by standing waves in the Sun (Thomson *et al* 1995), and is more generally evidenced in the very high temperatures of stellar coronae. Mass-loss in a wind provides a third mode for disposing of a radial momentum flux, that inherently maintains symmetry.

Our conclusions are premised on the supposition that the majority of the contraction momentum flux generated in the core is exported, whereas it is certainly reasonable to expect that in the absence of any other mode of disposing of it, it is thermally cancelled. This is the usual if little explored expectation. So the central question posed by the present exercise is the proper treatment to accord the contraction momentum flux.

Since the flux originates from a mechanical response of the stellar structure, it implicitly has a dynamical time scale. Its complete cancellation at the center, however, would associate it in most stellar contexts with the much longer photon diffusion timescale: these differing timescales can only be reconciled by the adjustment of the equilibrium thermal gradient around the center to the needs of cancellation, together with a perpetual mechanical oscillation of the whole structure. Yet this class of solution immediately requires the momentum flux to flow repeatedly through the center. Unless the momentum flux is immediately and completely cancelled at the center on its first arrival there, and there are no efficient mechanisms in any star for accomplishing this, its passage through the center to become in turn an outwardly directed radial flux mandates its transmission to the rest of the stellar structure, so that every other mode of disposing of a momentum flow needs exploration. The most obvious are (i) the consumption of the momentum flux in work against gravity, the inverse of the process from which it originated, and (ii) its export from the entire structure as a wind.

It is worth pondering further on the situation posed by an AGB star when contraction is not deemed to be the source of the momentum flux driving its wind. In that case the wind's flux must originate from some shell concentric with the center that lies beyond the core. And since action and reaction are equal and opposite, the launching of any momentum towards the outside implies the simultaneous and equal launching of a momentum flow towards the center. But the basic structure of a star is set by hydrostatic equilibrium, the more so as its center is approached, so the inward flow can neither be cancelled, nor absorbed by doing work against gravity, and so should reach the center. For consistency in this context, the inward flow has to be completely cancelled by the mechanism held to dispose of the central contraction momentum flux. Acceptance of this scenario therefore requires (i) an efficient, as yet unrecognized, central momentum cancellation mechanism; (ii) a mechanism for generating a radially directed momentum flow from a larger radial zone in the star; (iii) additional mechanisms for explaining large scale features such as bipolar flows (in AGB stars appeal is currently made to interaction with a close companion star, despite their rarity); (iv) auxiliary mechanisms for explaining winds from stars and proto-stars; (v) auxiliary hypotheses for explaining the large scale structural morphology of other contraction contexts. The need for an array of mechanisms is resolved by one application of Occam's razor. It is surely more reasonable to expect most contexts in which mass loss occurs in propinquity to gravitational contraction to be influenced by the inescapable generation of a contraction momentum flux there, than to invoke an array of mechanisms to replace it. My particular conclusion is that the ultimate source of the momentum flux powering every slow AGB wind is the momentum flux generated by its contracting core.

An early crude version of this paper was prepared while I was on leave at the Max Planck Institute für Radioastronomie; I thank Professor Wielebinski and Dr. Tom Wilson for their hospitality and a relaxed environment. This work is supported by the National Astronomy and Ionosphere Center, which is operated by Cornell University under a management agreement with the National Science Foundation.

## Appendix A

If the whole of the potential energy change of a homogeneous sphere under the addition of a mass, $\Delta M$, is available to increase its kinetic energy at the moment of addition, then

$$\Delta \Omega = \frac{3}{5} \frac{G M^2}{R_S} - \frac{3}{5} \frac{G(M + \Delta M)^2}{R_S} = -2 \Omega \left( \frac{\Delta M}{M} \right).$$

When $\rho \rightarrow \rho + \Delta \rho$ under this change, the kinetic energy of a spherical shell of mass $dm = 4 \pi r^2 \rho \, dr$ and radius $r$ in circular orbit about a central mass, $M(r)$, becomes

$$T_{shell} = \frac{1}{2} dm \, v_{rot}^2(r) + dm \left[ 2 |\Omega| \Delta M(r) / M(r) \right] = \frac{1}{2} dm \, v_{rot}^2(r) \left[ 1 + 4 \left( \Delta M(r) / M(r) \right) \right],$$

so the change in the velocity of $dm$ as a result of adding $\Delta M(r)$ can be thought of as

$$\Delta v(r) = 2 v_{rot}(r) \left( \Delta M(r) / M(r) \right) \quad \text{(A1)}.$$

The resultant change in $P$, the scalar sum of the momentum content of the homogeneous sphere defined by equation 11, when $\Delta M(r) = 4 \pi r^3 \Delta \rho / 3$, is

$$\begin{aligned}
\Delta P &= \int_o^{R_S} 2 v_{rot}(r) \left( \Delta M(r) / M(r) \right) 4 \pi r^2 \rho \, dr \\
&= (\Delta M / M) \int_o^{R_S} 2 v_{rot}(r) 4 \pi r^2 \rho \, dr \\
&= (\Delta M / M) \int_o^{R_S} 2 \left( \frac{G}{r} \frac{4 \pi r^3 \rho}{3} \right)^{1/2} 4 \pi r^2 \rho \, dr \\
&= (\Delta M / M) \int_o^{R_S} 2 \left( \frac{4 \pi \rho G}{3} \right)^{1/2} 4 \pi r^3 \rho \, dr \\
&= (\Delta M / M) \left[ \frac{2 * 3}{4} \left( \frac{4 \pi \rho G r^3}{3 r} \right)^{1/2} \left( \frac{4 \pi r^3 \rho}{3} \right) \right]_o^{R_S},
\end{aligned}$$

so from (11) & (13), $\quad \Delta P = 2 P (\Delta M / M) \qquad \qquad \qquad \qquad \qquad$ (A2)

Table 1: change in the contraction momentum flux from a 3/2 polytrope with luminosity

| $L_*$ ($L_o$) | $M_c$ ($M_o$) | $v_{esc}$ (km s$^{-1}$) | $\varepsilon$ | $\dot{P}$ ($L_*/c$) | $\dot{m}$ (lim)‡ ($10^{-6}$ $M_o$ yr$^{-1}$) |
|---|---|---|---|---|---|
| 4340  | 0.539 | 3910 | 0.891 | 1.431  | 2.4   |
| 5000  | 0.552 | 3983 | 0.893 | 1.478  | 3.0   |
| 7500  | 0.600 | 4259 | 0.899 | 1.666  | 5.6   |
| 10000 | 0.648 | 4539 | 0.905 | 1.876  | 9.0   |
| 15000 | 0.744 | 5116 | 0.914 | 2.386  | 19.0  |
| 20000 | 0.841 | 5730 | 0.922 | 3.070  | 35.0  |
| 30000 | 1.033 | 7169 | 0.935 | 5.506  | 104.2 |
| 40000 | 1.225 | 9287 | 0.946 | 12.898 | 349.6 |

‡ this upper limit to $\dot{m}$ is obtained by equating $\dot{P}$ to three times quasistatic limit of (35), with $T_{eff}$ = 1500 K

Table 2:

sensitivity of the momentum flux needs of photospheric mass-loss to $T_{eff}$

| $T_{eff}$ | $R_S$ | $v_{esc,p}$ | $r_d$ | $\dot{P}/10^6\,\dot{m}$ ‡ |
|---|---|---|---|---|
| (K) | ($R_\odot$) | km s$^{-1}$ | ($R_S$) | |
| 3000 | 244.4 | 39.5 | 15.59 | 5.82 |
| 2500 | 351.9 | 32.9 | 9.88 | 4.49 |
| 2000 | 549.8 | 26.3 | 5.66 | 3.14 |
| 1750 | 718.1 | 23.0 | 4.05 | 2.45 |
| 1500 | 977.4 | 19.8 | 2.76 | 1.75 |

‡ the $\dot{P}/\dot{m}$ ratio is calculated from the quasistatic bound of equation 35 for a 1 M$_\odot$ star with $L = 3440$ L$_\odot$ and n = 3/2.

**References**

Applegate, J. H. 1988, ApJ, 329, 803

Boothroyd, A. I., & Sackman, I. -J., 1992, ApJ 393, L21

Bowen, G. H., 1988, in *Pulsation and Mass Loss in Stars,* eds. R. Stalio and L. A. Wilson, (Kluwer) p3

Cassinelli, J. P., & MacGregor, K. B., 1986, *Physics of the Sun,* Ed. P. A. Sturrock (Boston: Reidel) v3, p47

Chandrasekhar, S. 1939, *Stellar Structure,* (Dover) p428

Dyck, C. H., Lockwood, G. W., & Capps, R. W., 1974, ApJ, 189, 89

Gilman, R. C., 1972, ApJ, 178, 423

Habing, H. J., Tignon, J., & Tielens, A. G. G. M., 1994, A&A, 286, 523

Hansen, C. J., & Kawaler, S. D., 1994, *"Stellar Interiors",* (Springer-Verlag: New York)

Hearn, A. 1990, in *"From Miras to Planetary Nebulae: Which Path for Stellar Evolution"* ed. M. O. Mennessier and A. Omont (Editiones Frontieres;Paris) p121

Holzer, T. E., 1976, in *Solar System Plasma Physics* ed. C. F. Kennel, L. J. Lanzerotti, & E. N. Parker (North Holland: Amsterdam) v1, p101

Hrivnak, B. J., 1996 (private communication)

Humphreys, E. M. L., Gray, M. D., Yates, J. A., Field, D., Bowen, G., & Diamond, P. J., 1996, MNRAS, 282, 1359

Iben, I. 1987, in *"Late Stages of Stellar Evolution",* ed. S. Kwok and S.R.Pottasch (Reidel; Dordrecht) p175

Ivezic, Z., & Elitzur, M., 1995, ApJ, 445, 415

Jura, M, 1984, ApJ, 282, 200

Knapp, G. R. 1986, ApJ, 311, 731

Knapp, G. R., & Morris, M., 1985, ApJ, 292, 640

Knapp, G. R., Phillips, T. G., Leighton, R. B., Lo, K. Y., Wannier, P. G., Wootten, H. A., & Huggins, P. J., 1982, ApJ, 252, 616

Kwok, S., 1975, ApJ, 198, 583

Kwok, S., 1987, Physics Rep, 156, 111

Lada, C. J., and Shu, F. H., 1990, Science., 248, 564

Landau, L. D., & Lifshitz, E. M., 1979, *"Fluid Mechanics"*, v6, p265

Lepine, J. R. D., Ortiz, R., & Epchtein, N., 1995, A&A, 299, 453

Le Sidaner, P., & Le Bertre, T., 1996, A&A, 314, 896

Lewis, B. M., Eder, J., and Terzian, Y. 1990, ApJ, 362, 634

Lewis, B. M., 1991, AJ, 101, 254

Lewis, B. M., 1987, in *Late Stages of Stellar Evolution,* ed. S. Kwok & S. Pottasch (Reidel, Dordrecht) p75

Lockwood, G. W., 1985, ApJS, 58, 167

Netzur, N, & Elitzur, M., 1993, ApJ, 410, 701

Olofsson. H., Eriksson, K., Gustafsson, B., & Carlstrom, U., 1993, ApJS, 87, 267

Parker, E. N., 1958, ApJ, 128, 664

Salpeter, E. E. 1974, ApJ, 193, 585

Thomson, D. J., Maclennan, C. G., & Lanzerotti, L. J., 1995, Nature, 376, 139

Whitworth, A. P. 1989, MNRAS, 236, 505

Wood, P. R. 1990, in *"From Miras to Planetary Nebulae: Which Path for Stellar Evolution"* ed. M. O. Mennessier and A. Omont (Editiones Frontieres;Paris) p67

Wood, P. R., Whiteoak, J. B., Hughes, S. M. G., Bessel, M. S., Gardner, F. F., & Hyland, H. H., 1992, ApJ, 397, 552

Wood, P. R., & Zarro, D., 1981, ApJ, 247, 247

Zuckerman, B., & Aller, L. H., 1986, ApJ, 301, 772

Legends:

Figure 1:   The galactic latitude distribution of IRAS sources with detected OH masers plotted against (25-12) μm   IR color index.  The thickest shells from the largest mass-loss rates appear on the right-hand side of the diagram, where they exhibit a notably smaller latitude distribution.

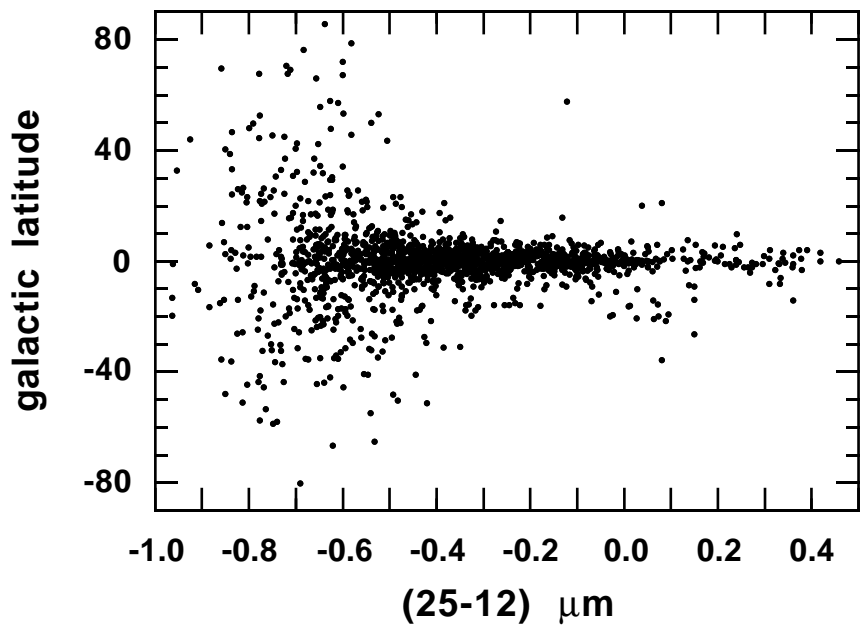